\documentclass[aps,pra,twocolumn,showpacs,floatfix]{revtex4}
\usepackage{graphicx}
\begin{document}
\title{Production of ultracold NH molecules by sympathetic cooling with Mg}
\author{Alisdair O. G. Wallis}
\affiliation{Department of Chemistry, Durham University, South
Road, Durham, DH1~3LE, United Kingdom}
\author{Jeremy M. Hutson}
\affiliation{Department of Chemistry, Durham University, South
Road, Durham, DH1~3LE, United Kingdom}

\date{\today}

\begin{abstract}
We carry out calculations on $M$-changing collisions of NH
($^3\Sigma^-$) molecules in magnetically trappable states using
a recently calculated potential energy surface. We show that
elastic collision rates are much faster than inelastic rates
for a wide range of fields at temperatures up to 10 mK and that
the ratio increases for lower temperatures and magnetic fields.
If NH molecules can be cooled to temperatures approaching 10 mK
and brought into contact with laser-cooled Mg then there is a
good prospect that sympathetic cooling can be achieved.
\end{abstract}


\maketitle



There is great interest in the production of samples of cold
molecules, at temperatures below 1 K, and ultracold molecules,
at temperatures below 1 mK. Ultracold molecules have many
potential applications in areas ranging from precision
measurement to quantum computing. They also offer new
possibilities for quantum control and controlled ultracold
chemistry.

There have been considerable successes in producing ultracold
molecules in laser-cooled atomic gases \cite{Hutson:IRPC:2006},
both by photoassociation \cite{Jones:RMP:2006} and by
magnetoassociation \cite{Kohler:RMP:2006}. Ni {\em et al.}\
\cite{Ni:KRb:2008} have very recently succeeded in producing
ultracold KRb in its ground rovibrational state by
magnetoassociation followed by stimulated Raman adiabatic
passage (STIRAP). However, such methods are limited to
molecules formed from atoms that can be laser-cooled, such as
the alkali metals. A wider range of molecules can be cooled
directly from high temperature to the millikelvin regime, using
methods such as buffer-gas cooling \cite{Weinstein:CaH:1998}
and Stark deceleration \cite{Bethlem:IRPC:2003}.
Low-field-seeking states of these cold molecules can then be
confined in electrostatic and magnetic traps. However, at
present the lowest temperature that can be achieved for
directly cooled molecules in static traps is around 10 mK. The
major challenge in this field is to find ways to cool such
molecules to the ultracold regime.

One of the most promising proposals for second-stage cooling is
{\em sympathetic cooling}, in which molecules are brought into
contact with a laser-cooled atomic gas that is already
ultracold. The hope is that thermalisation will occur to
produce ultracold molecules. However, atom-molecule potential
energy surfaces are often strongly anisotropic and the
anisotropy may drive fast inelastic collisions (relaxation).
Such collisions may prevent sympathetic cooling, because they
release kinetic energy and cause trap loss. A commonly stated
rule of thumb is that elastic collisions must be at least a
factor of 100 faster than inelastic collisions if sympathetic
cooling is to succeed.

In previous work, we have used high-level electronic structure
calculations to calculate interaction potentials for a variety
of systems that are candidates for sympathetic cooling. The
systems investigated include OH with Rb \cite{Lara:PRL:2006,
Lara:PRA:2007} and NH$_3$ and NH ($^3\Sigma$) with alkali-metal
and alkaline-earth atoms \cite{Zuchowski:NH3:2008,
Soldan:MgNH:2009}. Most of these systems were found to have
interaction potentials with deep wells and strong anisotropy,
and several of them also have ion-pair states that are expected
to cause additional inelasticity. It is unlikely that
sympathetic cooling would work for molecules in
low-field-seeking states in systems with high anisotropy.
However, Mg + NH and Be + NH were found to be much less
anisotropic and their ion-pair states are likely to be
energetically inaccessible in low-energy collisions
\cite{Soldan:MgNH:2009}. Cold NH in its ground $^3\Sigma$ state
can be cooled in a helium buffer gas and confined in a magnetic
trap \cite{Egorov:2004, Campbell:2007}, and laser cooling of Mg
to sub-Doppler temperatures has recently been achieved
\cite{Mehlstaubler:2008}. NH can also be decelerated and
trapped electrostatically in its excited $^1\Delta$ electronic
state \cite{vandeMeerakker:2006, Hoekstra:2007} and there is a
proposal to transfer the molecules to the ground state and
accumulate them in a magnetic trap \cite{vandeMeerakker:2001}.
The purpose of the present paper is therefore to use quantum
collision calculations to investigate whether sympathetic
cooling of magnetically trapped NH by Mg is a good prospect.

The energy levels of NH in a magnetic field are most
conveniently described using Hund's case (b), in which the
molecular rotation $n$ couples to the spin $s$ to produce a
total monomer angular momentum $j$. In zero field, each
rotational level $n$ is split into sub-levels labeled by $j$.
In a magnetic field, each sublevel splits further into $2j+1$
levels labeled by $m_j$, the projection of $j$ onto the axis
defined by the field. For the $n=0$ levels that are of most
interest for cold molecule studies, there is only a single
zero-field level with $j=1$ that splits into three components
with $m_j=+1$, 0 and $-1$. Molecules in the $m_j=+1$ state are
low-field-seeking and can be confined in a magnetic trap,
whereas those in the $m_j=0$ and $-1$ states are untrapped.

In the present paper, we carry out scattering calculations on
Mg + NH as a function of collision energy and magnetic field.
The calculations use the MOLSCAT package \cite{molscat:v14}, as
modified to handle collisions in magnetic fields
\cite{Gonzalez-Martinez:2007}. The collisions that are of most
interest are those of NH molecules that are initially in the
magnetically trappable $m_j=+1$ state, which may undergo
inelastic collisions to untrapped states with $m_j=0$ and $-1$.
The calculations are carried out in a partly coupled basis set
$|nsjm_j\rangle |LM_L\rangle$, where $L$ is the end-over-end
rotational angular momentum of the Mg atom and the NH molecule
about one another and $M_L$ is its projection on the axis
defined by the magnetic field. Hyperfine structure is
neglected. The matrix elements of the total Hamiltonian in this
basis are given in ref.\ \cite{Gonzalez-Martinez:2007}. The
only good quantum numbers during the collision are the parity
$p=(-1)^{n+L+1}$ and the total projection quantum number
$\mathcal{M} = m_j+M_L$; the total Hamiltonian is
block-diagonalized and scattering calculations are performed
separately for each parity and $\mathcal{M}$. The calculations
in the present work use basis sets with $n_{\rm max}=6$ and
$L_{\rm max}=8$.

MOLSCAT constructs a set of coupled equations in the
intermolecular distance $R$ and propagates them by the hybrid
log-derivative method of Alexander and Manolopoulos
\cite{Alexander:1987}, which uses a fixed-step-size
log-derivative propagator in the short-range region ($2.5 \leq
R < 50$~\AA) and a variable-step-size Airy propagator in the
long-range region ($50 \leq R \leq 250$~\AA). The
log-derivative solutions for each ${\mathcal M}$ are then
matched to asymptotic boundary conditions \cite{Johnson:1973}
to obtain the scattering matrix $S^{{\mathcal M}p}$ and
T-matrix $T^{{\mathcal M}p}=I-S^{{\mathcal M}p}$. It is useful
to decompose the integral cross sections between NH levels
($|\alpha\rangle =|n s j m_j\rangle)$ into sums of partial
cross sections characterized by the $L$ quantum number in the
incident channel,
\begin{equation}
\sigma^L_{\alpha \rightarrow \alpha' } =
\frac{\pi}{k_\alpha^2}\sum_{\mathcal{M}pL'}
|T^{\mathcal{M}p}_{\alpha ,LM_L\rightarrow\alpha',L'M_L'}|^2,
\end{equation}
where $M_L={\mathcal M}-m_j$, $M_L'={\mathcal M}-m_j'$,
$k_{\alpha}$ is the wave vector for incoming channel $\alpha$,
with collision energy $E = \hbar^2k_\alpha^2/2\mu$, and $\mu$
is the reduced mass of the colliding system. Since we focus
here on transitions among the $n=0$, $j=1$ levels, we
abbreviate the labels $\alpha$ to just $m_j$. For molecules
initially in the $m_j=+1$ state, the most important quantities
are the elastic and total inelastic cross sections; the latter
is the sum of the state-to-state inelastic cross sections to
the $m_j=0$ and $m_j=-1$ states.

The mechanism of spin relaxation for $^3\Sigma$ molecules has
been studied extensively \cite{Volpi:2002, Krems:henh:2003,
Krems:mfield:2004, Cybulski:2005, Campbell:2009}. At
sufficiently low energy, incoming channels with $L>0$ are
suppressed by centrifugal barriers. The heights of the barriers
are approximately $E^L_{\rm cf} = (\hbar L(L+1)/\mu)
^{\frac{3}{2}} (54C_6)^{\frac{1}{2}}$, which is 23 mK for Mg+NH
with $L=2$. For $L=0$ (s-wave scattering), $M_L=0$ and hence
$\mathcal{M}=m_j$. Since $\mathcal{M}$ is conserved there is no
outgoing channel with $L'=0$ for $m_j'=0$ or $-1$ and the
dominant relaxation channels for s-wave scattering have $L'=2$
in the outgoing channels.

The coupling between channels with different $m_j$ occurs via
the interplay of the spin-spin interaction and the potential
anisotropy. The spin-spin term in the NH Hamiltonian mixes the
$n=0$ and $n=2$ states with the same $j$ and $m_j$, and the
potential anisotropy then mixes states of different $L$ such
that $\Delta m_j + \Delta M_L = 0$.

In the absence of a magnetic field the thresholds for different
values of $m_j$ are degenerate. The presence of centrifugal
barriers in the outgoing channels strongly suppresses the
inelastic transitions, but the spin-relaxation cross section is
nevertheless nonzero at finite energy \cite{Krems:henh:2003,
Krems:mfield:2004}. Application of a magnetic field removes the
degeneracy, increasing the kinetic energy in the outgoing
channels and reducing the centrifugal suppression. By contrast,
the elastic cross section is dominated by $\Delta L=0$
collisions so are almost field-independent.

\begin{figure}
\includegraphics[width=0.95\linewidth]{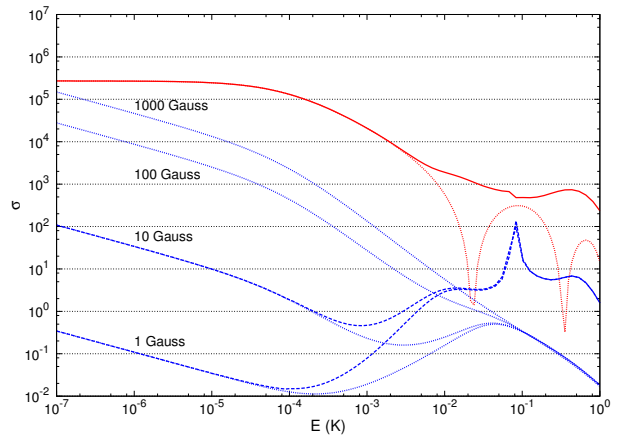}
\caption{(color online) Mg+NH elastic $\sigma^{L=0}_{m_j =
+1\rightarrow +1}$ (solid, red) and total inelastic
$\sigma^{L=0}_{m_j = +1\rightarrow 0}+\sigma^{L=0}_{m_j =
+1\rightarrow -1}$ (dashed, blue) s-wave cross sections as
a function of collision energy for various magnetic fields.
Integral cross sections including p, d and f waves ($L=1$, 2 and 3)
are also shown for the elastic cross section (red, dots)
and the total inelastic cross sections at 1 and 10 G (blue,
dots). This gives convergence for energies up to 100 mK.
All cross sections are in \AA$^2$.}\label{fig1}
\end{figure}

\begin{figure}
\includegraphics[width=0.95\linewidth]{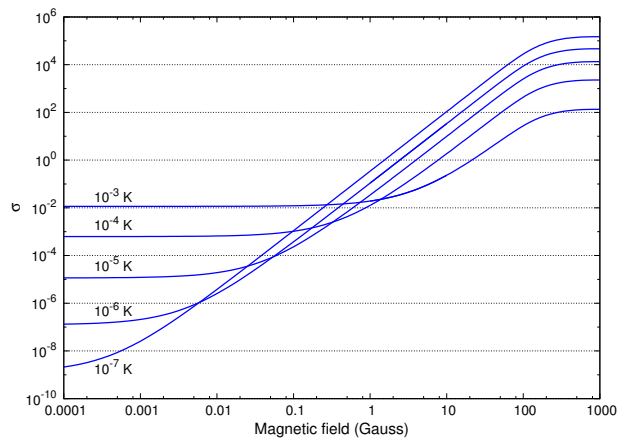}
\caption{(color online) Mg+NH total inelastic s-wave cross
sections $\sigma^{L=0}_{m_j = +1\rightarrow 0}+
\sigma^{L=0}_{m_j = +1\rightarrow -1}$ as a
function magnetic field strength for various collision
energies. All cross sections are in \AA$^2$.}\label{fig2}
\end{figure}

Fig.\ \ref{fig1} shows the s-wave elastic cross section
$\sigma^0_{\Delta m_j=0}$ and the total inelastic cross section
$\sigma^0_{\Delta m_j \neq 0}$ for initial $m_j=+1$ as a
function of energy for varying magnetic field strengths, while
Fig.\ \ref{fig2} shows the s-wave total inelastic cross
sections as a function of magnetic field for a number of
different collision energies. It may be seen that the total
inelastic cross section in the ultracold regime decreases
dramatically as the magnetic field is reduced.

For magnetic fields small enough that the NH Zeeman splitting
does not exceed the centrifugal barrier height in the outgoing
channel, the threshold behavior of the inelastic cross sections
as a function of energy and magnetic field can be understood as
described by Volpi and Bohn \cite{Volpi:2002}. Applying the
first-order distorted-wave Born approximation and approximating
the scattering wavefunction with the small-argument limit of
spherical Bessel functions, they obtained a simple
one-parameter formula,
\begin{equation}\label{eqnVBfit}
\sigma_{\alpha L \rightarrow \alpha' L'}(E,B) =
\sigma_{\alpha\alpha'}^{LL'} E^{L-\frac{1}{2}}
\left(E+\Delta m_j g \mu_0B\right)^{L'+\frac{1}{2}}
\end{equation}
where the factor $\sigma_{\alpha\alpha'}^{LL'}$ is independent
of energy and magnetic field and $\Delta m_j g \mu_0B$ is the
linear Zeeman shift where $\Delta m_j =m_j-m_j'$, $g$ is the
electron $g$-factor, and $\mu_0$ is the Bohr magneton. For the
centrifugal barrier to be exceeded in the dominant outgoing
channel ($m_j'=-1,L'= 2$) for a zero-energy collision requires
a magnetic field of 86 G. In figure~\ref{fig1} it can be seen
that the s-wave inelastic cross sections ($L=0\rightarrow
L'=2$) behave as (\ref{eqnVBfit}). When the collision energy is
less than the Zeeman shift, $E^{L-\frac{1}{2}}$ dominates and
the s-wave cross section is proportional to $E^{-\frac{1}{2}}$.
For higher collision energies, the second term in
(\ref{eqnVBfit}) also contributes and the s-wave cross sections
for low fields show $E^2$ behavior.

At low enough magnetic field, all the s-wave inelastic cross
sections in figure~\ref{fig2} flatten out to a zero-field value
proportional to $E^2$. At higher field they enter a region of
$B^{5/2}$ dependence. This continues until the centrifugal
barrier is exceeded in the outgoing channel and the cross
sections then flatten off with a value proportional to
$E^{-\frac{1}{2}}$.

As mentioned in the Introduction, the general rule of thumb for
sympathetic cooling to work is that the ratio $\gamma$ of
elastic to total inelastic cross sections must be greater than
about 100. It can see in figure~\ref{fig1} that for small
magnetic fields and low collision energies $\gamma$ is well in
excess of 100. However, at collision energies above $\sim
10^{-4}$ K, higher partial waves start contributing
significantly to the total cross sections. The total cross
sections incorporating additional p, d and f partial waves
($L=1$, 2 and 3) are included in Fig.\ \ref{fig1} for 1 and 10
G. There is a sharp peak in the d-wave inelastic cross section
around 75 mK, but everywhere else $\gamma$ remains in excess of
100 until partial waves with $L=4$ become important above 100
mK.

\begin{figure}
\includegraphics[width=0.92\linewidth]{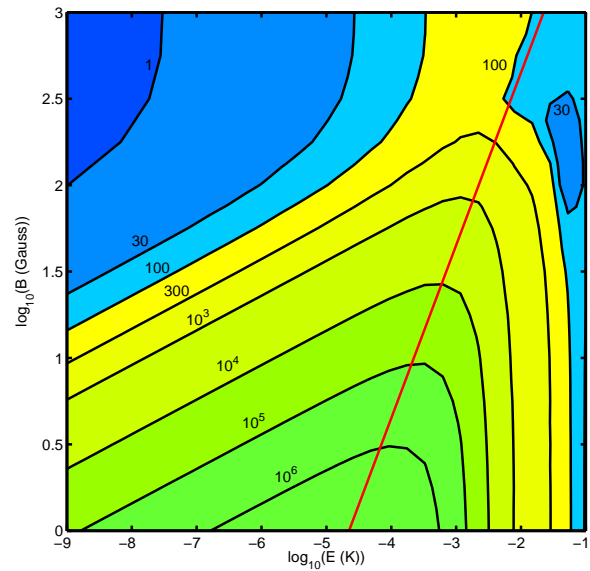}
\caption{(color online) Contour plot of the ratio $\gamma$ of
elastic to total inelastic cross section as a function of magnetic
field strength and collision energy. The red line shows the maximum
field sampled by trapped molecules in the $m_j=+1$ state
($B=6kT/g\mu_0$).}\label{fig3}
\end{figure}

In order to assess the prospects of sympathetic cooling for NH
with Mg, Fig.\ \ref{fig3} shows a contour plot of $\gamma$ as a
function of collision energy and magnetic field strength. At
the top left of this Figure, inelastic collisions are too fast
for sympathetic cooling to succeed. However, in an unbiased
magnetic trap with zero field at the center, trapped molecules
in $m_j=+1$ states at temperature $T$ will be distributed
according to a Boltzmann distribution with density $\rho$ given
by
\begin{equation}
\rho/\rho_0 = \exp\left(\frac{-m_jg\mu_0B}{k_BT}\right).
\end{equation}
At any given temperature on the energy axis of Fig.\
\ref{fig3}, only about 0.1\% of molecules will experience
fields greater than $B=6k_BT/g\mu_0$, which is shown as a red
line in Fig.\ \ref{fig3}. Most molecules in a magnetic trap
will experience fields below this line.

A precooled sample of molecules would initially have a
temperature on the order of tens of milliKelvin. For
temperatures up to about 10 mK, $\gamma$ is always greater than
100 in the thermally allowed region. Even above this, $\gamma$
remains close to 100. Nevertheless, it will be important to
precool the molecules as much as possible before sympathetic
cooling begins. As the sample is cooled towards sub-milliKelvin
temperatures, the maximum magnetic field strength sampled by
the molecules decreases, $\gamma$ increases and the trapped NH
molecules become {\em increasingly} stable to collisional
spin-relaxation. For an unbiased trap, $\gamma > 10^5$ at
temperatures below 0.1 mK. However, the addition of a bias
field to maintain the $m_j$ quantum number would limit the
increase in $\gamma$.

\begin{figure}
\includegraphics[width=0.95\linewidth]{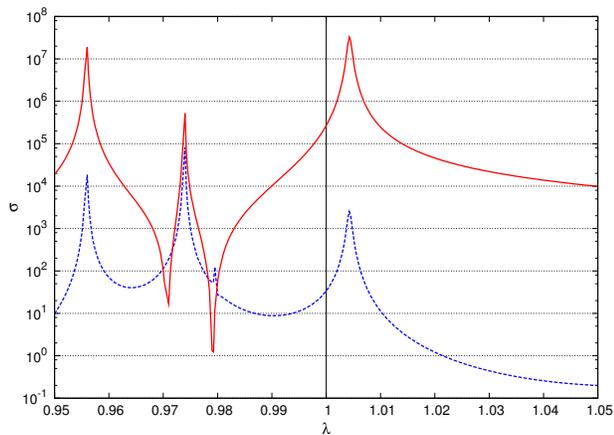}
\caption{(color online) S-wave elastic (solid, red) and total
inelastic (dashed, blue) cross sections as a function of the
potential scaling factor $\lambda$, calculated for a collision energy of 10$^{-6}$K
at a magnetic field of 10 Gauss. All cross sections are in
\AA$^2$.}\label{fig4}
\end{figure}

Scattering at low energy depends strongly on the details of the
potential energy surface. The potential energy surface used in
the present work is probably accurate to about 5\%. To explore
whether uncertainty in the potential surface affects our
conclusions, we consider the effect of a scaling factor
$\lambda$ that produces a modified potential energy surface
$V^{\rm scaled}(R,\theta) = \lambda V(R,\theta)$. The s-wave
elastic and total inelastic cross sections are shown as a
function of $\lambda$ in Fig.\ \ref{fig4} for a collision
energy of 1 $\mu$K at a field of 10 G. Both the elastic and
inelastic cross sections show strong resonance structures as
Mg-NH bound and quasi-bound states cross the low-field-seeking
threshold as the potential is varied. However, away from the
strong resonant structures the ratio of elastic to inelastic
cross sections remains large. This indicates that the our
conclusions are reasonably independent of the details of the
potential energy surface and confirms that Mg is a good
candidate for sympathetic cooling of magnetically trapped NH.

In conclusion, we have carried out calculations on
spin-changing collisions for NH($^3\Sigma^-$) molecules
colliding with Mg atoms. We find that the ratio of elastic to
inelastic cross sections exceeds 100, the factor required for
sympathetic cooling to succeed, for a wide range of collision
energies and magnetic fields. If precooled NH molecules at a
temperature around 10 mK can be brought into contact with
laser-cooled Mg, there is a good prospect that sympathetic
cooling will succeed. The inelastic losses decrease even
further as the temperature decreases, so that once sympathetic
cooling begins it will continue.


The authors are grateful to EPSRC for a research studentship
for AOGW and for funding the collaborative project CoPoMol
under the ESF EUROCORES programme EuroQUAM.

\bibliography{../../all}

\begin{thebibliography}{25}
\expandafter\ifx\csname natexlab\endcsname\relax\def\natexlab#1{#1}\fi
\expandafter\ifx\csname bibnamefont\endcsname\relax
  \def\bibnamefont#1{#1}\fi
\expandafter\ifx\csname bibfnamefont\endcsname\relax
  \def\bibfnamefont#1{#1}\fi
\expandafter\ifx\csname citenamefont\endcsname\relax
  \def\citenamefont#1{#1}\fi
\expandafter\ifx\csname url\endcsname\relax
  \def\url#1{\texttt{#1}}\fi
\expandafter\ifx\csname urlprefix\endcsname\relax\def\urlprefix{URL }\fi
\providecommand{\bibinfo}[2]{#2}
\providecommand{\eprint}[2][]{\url{#2}}

\bibitem[{\citenamefont{Hutson and Sold\'{a}n}(2006)}]{Hutson:IRPC:2006}
\bibinfo{author}{\bibfnamefont{J.~M.} \bibnamefont{Hutson}} \bibnamefont{and}
  \bibinfo{author}{\bibfnamefont{P.}~\bibnamefont{Sold\'{a}n}},
  \bibinfo{journal}{Int. Rev. Phys. Chem.} \textbf{\bibinfo{volume}{25}},
  \bibinfo{pages}{497} (\bibinfo{year}{2006}).

\bibitem[{\citenamefont{Jones et~al.}(2006)\citenamefont{Jones, Tiesinga, Lett,
  and Julienne}}]{Jones:RMP:2006}
\bibinfo{author}{\bibfnamefont{K.~M.} \bibnamefont{Jones}},
  \bibinfo{author}{\bibfnamefont{E.}~\bibnamefont{Tiesinga}},
  \bibinfo{author}{\bibfnamefont{P.~D.} \bibnamefont{Lett}}, \bibnamefont{and}
  \bibinfo{author}{\bibfnamefont{P.~S.} \bibnamefont{Julienne}},
  \bibinfo{journal}{Rev. Mod. Phys.} \textbf{\bibinfo{volume}{78}},
  \bibinfo{pages}{483} (\bibinfo{year}{2006}).

\bibitem[{\citenamefont{K\"{o}hler et~al.}(2006)\citenamefont{K\"{o}hler,
  Goral, and Julienne}}]{Kohler:RMP:2006}
\bibinfo{author}{\bibfnamefont{T.}~\bibnamefont{K\"{o}hler}},
  \bibinfo{author}{\bibfnamefont{K.}~\bibnamefont{Goral}}, \bibnamefont{and}
  \bibinfo{author}{\bibfnamefont{P.~S.} \bibnamefont{Julienne}},
  \bibinfo{journal}{Rev. Mod. Phys.} \textbf{\bibinfo{volume}{78}},
  \bibinfo{pages}{1311} (\bibinfo{year}{2006}).

\bibitem[{\citenamefont{Ni et~al.}(2008)\citenamefont{Ni, Ospelkaus, {de
  Miranda}, Pe'er, Neyenhuis, Zirbel, Kotochigova, Julienne, Jin, and
  Ye}}]{Ni:KRb:2008}
\bibinfo{author}{\bibfnamefont{K.-K.} \bibnamefont{Ni}},
  \bibinfo{author}{\bibfnamefont{S.}~\bibnamefont{Ospelkaus}},
  \bibinfo{author}{\bibfnamefont{M.~H.~G.} \bibnamefont{{de Miranda}}},
  \bibinfo{author}{\bibfnamefont{A.}~\bibnamefont{Pe'er}},
  \bibinfo{author}{\bibfnamefont{B.}~\bibnamefont{Neyenhuis}},
  \bibinfo{author}{\bibfnamefont{J.~J.} \bibnamefont{Zirbel}},
  \bibinfo{author}{\bibfnamefont{S.}~\bibnamefont{Kotochigova}},
  \bibinfo{author}{\bibfnamefont{P.~S.} \bibnamefont{Julienne}},
  \bibinfo{author}{\bibfnamefont{D.~S.} \bibnamefont{Jin}}, \bibnamefont{and}
  \bibinfo{author}{\bibfnamefont{J.}~\bibnamefont{Ye}},
  \bibinfo{journal}{Science} \textbf{\bibinfo{volume}{322}},
  \bibinfo{pages}{231} (\bibinfo{year}{2008}).

\bibitem[{\citenamefont{Weinstein et~al.}(1998)\citenamefont{Weinstein,
  deCarvalho, Guillet, Friedrich, and Doyle}}]{Weinstein:CaH:1998}
\bibinfo{author}{\bibfnamefont{J.~D.} \bibnamefont{Weinstein}},
  \bibinfo{author}{\bibfnamefont{R.}~\bibnamefont{deCarvalho}},
  \bibinfo{author}{\bibfnamefont{T.}~\bibnamefont{Guillet}},
  \bibinfo{author}{\bibfnamefont{B.}~\bibnamefont{Friedrich}},
  \bibnamefont{and} \bibinfo{author}{\bibfnamefont{J.~M.} \bibnamefont{Doyle}},
  \bibinfo{journal}{Nature} \textbf{\bibinfo{volume}{395}},
  \bibinfo{pages}{148} (\bibinfo{year}{1998}).

\bibitem[{\citenamefont{Bethlem and Meijer}(2003)}]{Bethlem:IRPC:2003}
\bibinfo{author}{\bibfnamefont{H.~L.} \bibnamefont{Bethlem}} \bibnamefont{and}
  \bibinfo{author}{\bibfnamefont{G.}~\bibnamefont{Meijer}},
  \bibinfo{journal}{Int. Rev. Phys. Chem.} \textbf{\bibinfo{volume}{22}},
  \bibinfo{pages}{73} (\bibinfo{year}{2003}).

\bibitem[{\citenamefont{Lara et~al.}(2006)\citenamefont{Lara, Bohn, Potter,
  Sold\'{a}n, and Hutson}}]{Lara:PRL:2006}
\bibinfo{author}{\bibfnamefont{M.}~\bibnamefont{Lara}},
  \bibinfo{author}{\bibfnamefont{J.~L.} \bibnamefont{Bohn}},
  \bibinfo{author}{\bibfnamefont{D.}~\bibnamefont{Potter}},
  \bibinfo{author}{\bibfnamefont{P.}~\bibnamefont{Sold\'{a}n}},
  \bibnamefont{and} \bibinfo{author}{\bibfnamefont{J.~M.}
  \bibnamefont{Hutson}}, \bibinfo{journal}{Phys. Rev. Lett.}
  \textbf{\bibinfo{volume}{97}}, \bibinfo{pages}{183201}
  (\bibinfo{year}{2006}).

\bibitem[{\citenamefont{Lara et~al.}(2007)\citenamefont{Lara, Bohn, Potter,
  Sold\'{a}n, and Hutson}}]{Lara:PRA:2007}
\bibinfo{author}{\bibfnamefont{M.}~\bibnamefont{Lara}},
  \bibinfo{author}{\bibfnamefont{J.~L.} \bibnamefont{Bohn}},
  \bibinfo{author}{\bibfnamefont{D.~E.} \bibnamefont{Potter}},
  \bibinfo{author}{\bibfnamefont{P.}~\bibnamefont{Sold\'{a}n}},
  \bibnamefont{and} \bibinfo{author}{\bibfnamefont{J.~M.}
  \bibnamefont{Hutson}}, \bibinfo{journal}{Phys. Rev. A}
  \textbf{\bibinfo{volume}{75}}, \bibinfo{pages}{012704}
  (\bibinfo{year}{2007}).

\bibitem[{\citenamefont{\.Zuchowski and Hutson}(2008)}]{Zuchowski:NH3:2008}
\bibinfo{author}{\bibfnamefont{P.~S.} \bibnamefont{\.Zuchowski}}
  \bibnamefont{and} \bibinfo{author}{\bibfnamefont{J.~M.}
  \bibnamefont{Hutson}}, \bibinfo{journal}{Phys. Rev. A}
  \textbf{\bibinfo{volume}{78}}, \bibinfo{pages}{022701}
  (\bibinfo{year}{2008}).

\bibitem[{\citenamefont{Sold\'{a}n et~al.}(2009)\citenamefont{Sold\'{a}n,
  \.Zuchowski, and Hutson}}]{Soldan:MgNH:2009}
\bibinfo{author}{\bibfnamefont{P.}~\bibnamefont{Sold\'{a}n}},
  \bibinfo{author}{\bibfnamefont{P.~S.} \bibnamefont{\.Zuchowski}},
  \bibnamefont{and} \bibinfo{author}{\bibfnamefont{J.~M.}
  \bibnamefont{Hutson}}, \bibinfo{journal}{Faraday Discussions}
  \textbf{\bibinfo{volume}{142}}, \bibinfo{pages}{paper 6}
  (\bibinfo{year}{2009}).

\bibitem[{\citenamefont{Egorov et~al.}(2004)\citenamefont{Egorov, Campbell,
  Friedrich, Maxwell, Tsikata, {van Buuren}, and Doyle}}]{Egorov:2004}
\bibinfo{author}{\bibfnamefont{D.}~\bibnamefont{Egorov}},
  \bibinfo{author}{\bibfnamefont{W.~C.} \bibnamefont{Campbell}},
  \bibinfo{author}{\bibfnamefont{B.}~\bibnamefont{Friedrich}},
  \bibinfo{author}{\bibfnamefont{S.~E.} \bibnamefont{Maxwell}},
  \bibinfo{author}{\bibfnamefont{E.}~\bibnamefont{Tsikata}},
  \bibinfo{author}{\bibfnamefont{L.~D.} \bibnamefont{{van Buuren}}},
  \bibnamefont{and} \bibinfo{author}{\bibfnamefont{J.~M.} \bibnamefont{Doyle}},
  \bibinfo{journal}{Eur. Phys. J. D} \textbf{\bibinfo{volume}{31}},
  \bibinfo{pages}{307} (\bibinfo{year}{2004}).

\bibitem[{\citenamefont{Campbell et~al.}(2007)\citenamefont{Campbell, Tsikata,
  Lu, van Buuren, and Doyle}}]{Campbell:2007}
\bibinfo{author}{\bibfnamefont{W.~C.} \bibnamefont{Campbell}},
  \bibinfo{author}{\bibfnamefont{E.}~\bibnamefont{Tsikata}},
  \bibinfo{author}{\bibfnamefont{H.-I.} \bibnamefont{Lu}},
  \bibinfo{author}{\bibfnamefont{L.~D.} \bibnamefont{van Buuren}},
  \bibnamefont{and} \bibinfo{author}{\bibfnamefont{J.~M.} \bibnamefont{Doyle}},
  \bibinfo{journal}{Phys. Rev. Lett.} \textbf{\bibinfo{volume}{98}},
  \bibinfo{pages}{213001} (\bibinfo{year}{2007}).

\bibitem[{\citenamefont{Mehlst\"aubler
  et~al.}(2008)\citenamefont{Mehlst\"aubler, Moldenhauer, Riedmann, Rehbein,
  Friebe, Rasel, and Ertmer}}]{Mehlstaubler:2008}
\bibinfo{author}{\bibfnamefont{T.~E.} \bibnamefont{Mehlst\"aubler}},
  \bibinfo{author}{\bibfnamefont{K.}~\bibnamefont{Moldenhauer}},
  \bibinfo{author}{\bibfnamefont{M.}~\bibnamefont{Riedmann}},
  \bibinfo{author}{\bibfnamefont{N.}~\bibnamefont{Rehbein}},
  \bibinfo{author}{\bibfnamefont{J.}~\bibnamefont{Friebe}},
  \bibinfo{author}{\bibfnamefont{E.~M.} \bibnamefont{Rasel}}, \bibnamefont{and}
  \bibinfo{author}{\bibfnamefont{W.}~\bibnamefont{Ertmer}},
  \bibinfo{journal}{Phys. Rev. A} \textbf{\bibinfo{volume}{77}},
  \bibinfo{pages}{021402R} (\bibinfo{year}{2008}).

\bibitem[{\citenamefont{{van de Meerakker} et~al.}(2006)\citenamefont{{van de
  Meerakker}, Labazan, Hoekstra, K\"upper, and Meijer}}]{vandeMeerakker:2006}
\bibinfo{author}{\bibfnamefont{S.~Y.~T.} \bibnamefont{{van de Meerakker}}},
  \bibinfo{author}{\bibfnamefont{I.}~\bibnamefont{Labazan}},
  \bibinfo{author}{\bibfnamefont{S.}~\bibnamefont{Hoekstra}},
  \bibinfo{author}{\bibfnamefont{J.}~\bibnamefont{K\"upper}}, \bibnamefont{and}
  \bibinfo{author}{\bibfnamefont{G.}~\bibnamefont{Meijer}},
  \bibinfo{journal}{J. Phys. B -- At. Mol. Opt. Phys.}
  \textbf{\bibinfo{volume}{39}}, \bibinfo{pages}{S1077} (\bibinfo{year}{2006}).

\bibitem[{\citenamefont{Hoekstra et~al.}(2007)\citenamefont{Hoekstra, Metsala,
  Zieger, Scharfenberg, Gilijamse, Meijer, and {van de
  Meerakker}}}]{Hoekstra:2007}
\bibinfo{author}{\bibfnamefont{S.}~\bibnamefont{Hoekstra}},
  \bibinfo{author}{\bibfnamefont{M.}~\bibnamefont{Metsala}},
  \bibinfo{author}{\bibfnamefont{P.~C.} \bibnamefont{Zieger}},
  \bibinfo{author}{\bibfnamefont{L.}~\bibnamefont{Scharfenberg}},
  \bibinfo{author}{\bibfnamefont{J.~J.} \bibnamefont{Gilijamse}},
  \bibinfo{author}{\bibfnamefont{G.}~\bibnamefont{Meijer}}, \bibnamefont{and}
  \bibinfo{author}{\bibfnamefont{S.~Y.~T.} \bibnamefont{{van de Meerakker}}},
  \bibinfo{journal}{Phys. Rev. A} \textbf{\bibinfo{volume}{76}},
  \bibinfo{pages}{063408} (\bibinfo{year}{2007}).

\bibitem[{\citenamefont{{van de Meerakker} et~al.}(2001)\citenamefont{{van de
  Meerakker}, Jongma, Bethlem, and Meijer}}]{vandeMeerakker:2001}
\bibinfo{author}{\bibfnamefont{S.~Y.~T.} \bibnamefont{{van de Meerakker}}},
  \bibinfo{author}{\bibfnamefont{R.~T.} \bibnamefont{Jongma}},
  \bibinfo{author}{\bibfnamefont{H.~L.} \bibnamefont{Bethlem}},
  \bibnamefont{and} \bibinfo{author}{\bibfnamefont{G.}~\bibnamefont{Meijer}},
  \bibinfo{journal}{Phys. Rev. A} \textbf{\bibinfo{volume}{64}},
  \bibinfo{pages}{041401} (\bibinfo{year}{2001}).

\bibitem[{\citenamefont{Hutson and Green}(1994)}]{molscat:v14}
\bibinfo{author}{\bibfnamefont{J.~M.} \bibnamefont{Hutson}} \bibnamefont{and}
  \bibinfo{author}{\bibfnamefont{S.}~\bibnamefont{Green}},
  \emph{\bibinfo{title}{{MOLSCAT} computer program, version 14}},
  \bibinfo{howpublished}{distributed by Collaborative Computational Project
  No.\ 6 of the UK Engineering and Physical Sciences Research Council}
  (\bibinfo{year}{1994}).

\bibitem[{\citenamefont{Gonz\'{a}lez-Mart\'{\i}nez and
  Hutson}(2007)}]{Gonzalez-Martinez:2007}
\bibinfo{author}{\bibfnamefont{M.~L.} \bibnamefont{Gonz\'{a}lez-Mart\'{\i}nez}}
  \bibnamefont{and} \bibinfo{author}{\bibfnamefont{J.~M.}
  \bibnamefont{Hutson}}, \bibinfo{journal}{Phys. Rev. A}
  \textbf{\bibinfo{volume}{75}}, \bibinfo{pages}{022702}
  (\bibinfo{year}{2007}).

\bibitem[{\citenamefont{Alexander and Manolopoulos}(1987)}]{Alexander:1987}
\bibinfo{author}{\bibfnamefont{M.~H.} \bibnamefont{Alexander}}
  \bibnamefont{and} \bibinfo{author}{\bibfnamefont{D.~E.}
  \bibnamefont{Manolopoulos}}, \bibinfo{journal}{J. Chem. Phys.}
  \textbf{\bibinfo{volume}{86}}, \bibinfo{pages}{2044} (\bibinfo{year}{1987}).

\bibitem[{\citenamefont{Johnson}(1973)}]{Johnson:1973}
\bibinfo{author}{\bibfnamefont{B.~R.} \bibnamefont{Johnson}},
  \bibinfo{journal}{J. Comput. Phys.} \textbf{\bibinfo{volume}{13}},
  \bibinfo{pages}{445} (\bibinfo{year}{1973}).

\bibitem[{\citenamefont{Volpi and Bohn}(2002)}]{Volpi:2002}
\bibinfo{author}{\bibfnamefont{A.}~\bibnamefont{Volpi}} \bibnamefont{and}
  \bibinfo{author}{\bibfnamefont{J.~L.} \bibnamefont{Bohn}},
  \bibinfo{journal}{Phys. Rev. A} \textbf{\bibinfo{volume}{65}},
  \bibinfo{pages}{052712} (\bibinfo{year}{2002}).

\bibitem[{\citenamefont{Krems et~al.}(2003)\citenamefont{Krems, Sadeghpour,
  Dalgarno, Zgid, K{\l}os, and Cha{\l}asi\'{n}ski}}]{Krems:henh:2003}
\bibinfo{author}{\bibfnamefont{R.~V.} \bibnamefont{Krems}},
  \bibinfo{author}{\bibfnamefont{H.~R.} \bibnamefont{Sadeghpour}},
  \bibinfo{author}{\bibfnamefont{A.}~\bibnamefont{Dalgarno}},
  \bibinfo{author}{\bibfnamefont{D.}~\bibnamefont{Zgid}},
  \bibinfo{author}{\bibfnamefont{J.}~\bibnamefont{K{\l}os}}, \bibnamefont{and}
  \bibinfo{author}{\bibfnamefont{G.}~\bibnamefont{Cha{\l}asi\'{n}ski}},
  \bibinfo{journal}{Phys. Rev. A} \textbf{\bibinfo{volume}{68}},
  \bibinfo{pages}{051401(R)} (\bibinfo{year}{2003}).

\bibitem[{\citenamefont{Krems and Dalgarno}(2004)}]{Krems:mfield:2004}
\bibinfo{author}{\bibfnamefont{R.~V.} \bibnamefont{Krems}} \bibnamefont{and}
  \bibinfo{author}{\bibfnamefont{A.}~\bibnamefont{Dalgarno}},
  \bibinfo{journal}{J. Chem. Phys.} \textbf{\bibinfo{volume}{120}},
  \bibinfo{pages}{2296} (\bibinfo{year}{2004}).

\bibitem[{\citenamefont{Cybulski et~al.}(2005)\citenamefont{Cybulski, Krems,
  Sadeghpour, Dalgarno, K{\l}os, Groenenboom, {van der Avoird}, Zgid, and
  Cha{\l}asi\'{n}ski}}]{Cybulski:2005}
\bibinfo{author}{\bibfnamefont{H.}~\bibnamefont{Cybulski}},
  \bibinfo{author}{\bibfnamefont{R.~V.} \bibnamefont{Krems}},
  \bibinfo{author}{\bibfnamefont{H.~R.} \bibnamefont{Sadeghpour}},
  \bibinfo{author}{\bibfnamefont{A.}~\bibnamefont{Dalgarno}},
  \bibinfo{author}{\bibfnamefont{J.}~\bibnamefont{K{\l}os}},
  \bibinfo{author}{\bibfnamefont{G.~C.} \bibnamefont{Groenenboom}},
  \bibinfo{author}{\bibfnamefont{A.}~\bibnamefont{{van der Avoird}}},
  \bibinfo{author}{\bibfnamefont{D.}~\bibnamefont{Zgid}}, \bibnamefont{and}
  \bibinfo{author}{\bibfnamefont{G.}~\bibnamefont{Cha{\l}asi\'{n}ski}},
  \bibinfo{journal}{J. Chem. Phys.} \textbf{\bibinfo{volume}{122}},
  \bibinfo{pages}{094307} (\bibinfo{year}{2005}).

\bibitem[{\citenamefont{Campbell et~al.}(2009)\citenamefont{Campbell,
  Tscherbul, Lu, Tsikata, Krems, and Doyle}}]{Campbell:2009}
\bibinfo{author}{\bibfnamefont{W.~C.} \bibnamefont{Campbell}},
  \bibinfo{author}{\bibfnamefont{T.~V.} \bibnamefont{Tscherbul}},
  \bibinfo{author}{\bibfnamefont{H.~I.} \bibnamefont{Lu}},
  \bibinfo{author}{\bibfnamefont{E.}~\bibnamefont{Tsikata}},
  \bibinfo{author}{\bibfnamefont{R.~V.} \bibnamefont{Krems}}, \bibnamefont{and}
  \bibinfo{author}{\bibfnamefont{J.~M.} \bibnamefont{Doyle}},
  \bibinfo{journal}{Phys. Rev. Lett.} \textbf{\bibinfo{volume}{102}},
  \bibinfo{pages}{013003} (\bibinfo{year}{2009}).

\end{thebibliography}

\end{document}